\documentclass[a4paper,11pt]{article}
\usepackage{jheppub} 
\usepackage[T1]{fontenc}
\usepackage{amssymb}
\usepackage{amsmath}
\usepackage{graphicx}
\usepackage{bm}
\usepackage{dcolumn}
\usepackage{amsfonts}%
\providecommand{\U}[1]{\protect\rule{.1in}{.1in}}
\providecommand{\U}[1]{\protect\rule{.1in}{.1in}}
\title{\boldmath Analytic results for the massive sunrise integral in the context of an
	alternative perturbative calculational method}

\author[a,1]{G. Dallabona,\note{Corresponding author.}}
\author[b]{ O. A. Battistel}

\affiliation[a]{Departamento de F\'{\i}sica, Universidade Federal de Lavras, Cx. Postal, 37203-022, Lavras, MG, Brazil.}
\affiliation[b]{Departamento de F\'{\i}sica, Universidade Federal de Santa Maria, 97119-900, Santa Maria, RS, Brazil.}

\emailAdd{gilson.dallabona@ufla.br}
\emailAdd{orimar.battistel@gmail.com}

\abstract{
An explicit investigation about the equal-mass two-loop sunrise Feynman graph
is performed. Such perturbative amplitude is related with many important
physical process treated in the standard model context. The background of this
investigation is an alternative strategy to handle with the divergences
typical of perturbative solutions of quantum field theory. Since its
proposition, the mentioned method was exhaustively used to calculate and
manipulate one-loop Feynman integrals with a great success. However, the great
advances in precision of experimental data collected in particle physics
colliders have pushed up theoretical physicists to improve their predictions
through multi-loops calculations. In the present job, we describe the main
steps required to perform two-loops calculations within the context of the
referred method. We show that the same rules used for one-loop calculations
are enough to deal with two-loops graphs as well. Analytic results for the
sunrise graph are obtained in terms of elliptic multiple polylogarithms as
well as a numerical analysis is provided.}


\begin{document}
\maketitle
	
\section{Introduction}
Many years have passed since the creation of quantum field theories (QFTs).
The amount of the experimental data produced in particle physics colliders, as
well as their accuracy, has been incredible increased along this time. The
newest and remarkable experimental machine, the larger hadron collider (LHC),
the most powerful one ever built, is the main responsible for this. From a lot
of data extracted from the performed experiments, it was possible to confirm
the existence of the Higgs particle (predicted many years before), the last
crucial ingredient of the standard model. It is expected that the experiments
are not restricted to confirm well-known theoretical predictions but also to
open new frontiers like occurred recently when the interpretation of
experimental results seems to give support, at least in hypothesis level, for
the existence of a new fundamental force of nature, fact that, if confirmed,
will represent a true revolution in the conception of the fundamental
particles and their interaction which will requires new theoretical
developments. This increasingly higher experimental precision has forced
theoretical particle physicists to improve its predictions as well. Since the
majority of predictions extracted from the theoretical models, at high energy,
are made within the framework of perturbative calculations, this, in practice,
imply to go beyond one-loop approximation in calculations involving Feynman
loop diagrams. This means to deal with the challenging goal of calculating
Feynman integrals with two or more loops.

As it is well known, perhaps the major issue involved in this kind of
calculation, is due to the fact that, in general, the Feynman integrals are
divergent mathematical objects. Then, in order to give a meaning for such
objects, one need to define a tool to deal with such kind of undefined
quantities. The most used techniques are based on regularization methods.
Essentially, a regularization consist in introducing some kind of modification
in the Feynman integrals, resulting in its convergence. However, the results
obtained, unfortunately, are regularization dependent.

Throughout QFT's history, many regularization methods were proposed and used
mainly to perform one-loop calculations. Among all these methods, only the
dimensional regularization (DR) \cite{Hooft, Bollini, Ashmore} showed
potentiality to be used in calculations beyond one-loop and the reason is more
of practical order than conceptual. Usually, in such regularizations, like
Pauli-Villars \cite{Pauli}, the number of parameters increase fast with the
number of loops, which often makes the integrals calculation unworkable or
produces results that are difficult to use or interpret.

By using DR as the background framework, many powerful and interesting
techniques has been developed in order to solve multi-loop/multi-legs
integrals, divergent or not. In spite that, even today there are open problems
about this issue, such that this a very active field of research (for some
reviews see refs. \cite{Buttaret, Weinzierl, Smirnov, Argeri, Henn}). In the
recent literature, the most popular and successful technique used to deal with
multi-loop Feynman diagrams consist, as a first step, in to reduce an
arbitrary scalar integral to a linear combination of a finite set of basis
integrals, the so called master integrals. This is done by applying the
integration-by-parts method \cite{Chetyrkin, Tkachov}, Lorentz-invariance
identities, and symmetry properties of these integrals. In its turn, the
master integrals are usually solved in two main ways. One of such approaches,
which is widely used, is given by the well-known differential equations method
\cite{Kotikov, Kotikov2, Kotikov3, Bern, Remiddi, Gehrmann}. In this method,
one can derive a system of differential equations in the kinematic invariants
satisfied by a set of master integrals, belonging to a certain topology, and
then try to solve it in order to obtain a solution for such integrals without
need a direct integration. Another possible technique start by introducing a
parametric transformation in the Feynman integrals, in order to allow the
integration over the loop momenta, for then to perform a direct integration
over the parameters of such transformation.

The results of Feynman integrals calculations are commonly written in terms of
special transcendental functions. For example, it was realized, many years
ago, that all the one-loop Feynman integrals can be solved in terms of
multiple polylogarithms (MPLs) functions \cite{Goncharov, Goncharov2,
Remiddi4, Vollinga}. However, for two-loop and beyond this cannot be done for
all topologies. The two-loop equal mass sunrise integral is the simplest
Feynman integral which is known that cannot be expressed in terms of MPLs
since it involves integrals of elliptic type. Thus, it was necessary to
consider generalizations of ordinary MPLs, called usually elliptic multiple
polylogarithms (eMPLs) \cite{Beilinson, Levin, Brown, Broedel, Broedel2,
Broedel3, Broedel4, Ablinger}. Along the last decade, various slightly different
definitions of such transcendental functions have been introduced in the
literature. Some recent definitions of eMPLs have been proposed in
Refs. \cite{Broedel, Broedel2}, where the authors present a rigorous and
detailed discussion of these special functions, as well as give examples of
applications such as the calculation of the sunset diagram. The present paper is in the same line of reasoning as these works and intends to add some contribution to this problem.

With the help of the cited methods, among others, much progress has been done
along the last decades with respect to the Feynman integrals calculations. In
spite of the success achieved so far, it would be desirable and useful, for
many reasons, to have at our disposal an alternative method to DR to deal with
multi-loop calculations in a consistent way and which can be easily
applicable. In fact, such a method already exist and was proposed in the early
20's by one of the present authors in his doctoral thesis \cite{Orimar-Thesis}%
. Since then, this alternative strategy was applied to many problems involving
one-loop calculations with very quite successful results (some examples can be
seen in the Refs. \cite{Orimar1, Orimar2, Orimar3, Orimar4, Orimar5, Fonseca,
Cherchiglia, Gnendiger, Fargnoli}). In such a papers, many subtle aspects of
the perturbative calculation were raised and then solved in a clear way. The
strategy is based on a very simple idea, which is to avoid, as much as
possible, to perform the integration of a purely divergent integral. Instead,
the method suggest to extract the physical content of a divergent integral by
rewritten its integrand in a such way that it can be splitted into a sum of
finite terms and purely divergent objects, which are the ones that do not have
physical parameters inside them. One way that this can be done is by adopting
a convenient representation for the propagators, as we will show in this
paper. The finite integrals can be integrated out without restrictions and the
set of divergent ones is reorganized into scalar objects and tensor surface
terms. By following this approach, the final results obtained retain all the
original properties of the integrals, making possible to make more general
analyses of the pertinent physical process. In many situations, this
represents an advantage with respect to the traditional regularization methods.

However, the potentiality of the method was little explored in scenarios
involving perturbative corrections at two-loops order or more. The few jobs
published in this line have just focused in some particular aspects, such as
massless diagrams \cite{Brizola, Carneiro, Dias, Cherchiglia2}, and so they
did not exploited the full potentiality of the method in the context
multi-loop calculations. In this paper, we will start to fulfill this gap by
using the method to calculate the well-known massive sunrise graph. This
Feynman two-loop diagram is very important, since it appears in many physical
process treated in the standard model context, and, therefore, have been
studied in several papers by means of different approaches \cite{Broadhurst,
Broadhurst2, Berends, Bauberger, Bauberger2, Post, Caffo, Berends2, Groote,
Groote2, Davydychev, Bashir, Caffo3, Onishchenko, Laporta, Tarasov, Pozzorini,
Bailey, Caffo2, Muller, Kniehl, Adams, Adams2, Remiddi2, Adams3, Bloch,
Adams4, Remiddi3, Bogner, Bogner2}. By handling this problem, we will discuss
the main steps in order to perform two-loop computations within the method. In
particular, we will see that the same recipe formulated to perform one-loop
calculations works for multi-loop calculations as well. The issue of
divergences present in subtopologies is also solved in a natural way.

In order to realize our program we organize the work as follows. In the
Section II, we briefly review the aforementioned method, highlighting the main
steps used to calculate some standard scalar one-loop integrals. The
calculations of the equal mass sunrise integral, through the method, are
explicitly shown in the Section III. As it will be shown in the prescription
stated in the Section II, the mass parameter used to define what we call basic
divergent objects is arbitrary. Thus, in Section IV, we obtain useful
relations which connect typical two-loop basic divergent objects defined at
two different mass scales. In the Section V, we perform explicitly the
integration over the Feynman parameters and write the results in terms of
eMPLs. A numerical analysis of the results obtained in the Section V is
discussed in the Section VI and, finally, the final comments and remarks are
given in the Section VII.

\section{The method to handle (divergent) Feynman integrals - a brief review
\label{sec_method}}

All perturbative amplitudes, defining typical physical processes (scattering
and particle decays) in the framework of QFTs, are, in fact, reduced to a
combination of Feynman integrals. Unfortunately, many of them are divergent
structures, which require much care with their manipulations. In this section
we briefly describe the method that we adopt to treat divergent Feynman
integrals by reviewing how it works in calculations involving one-loop
integrals. The same prescription will be used to treat the two-loop sunrise
integral in the next section. Much of the material succinctly presented in
this section can be found extensively discussed in the Refs. \cite{Orimar4,
Orimar6}.

A $d$-dimensional one-loop scalar $N$-point integral, with arbitrary internal
(loop) momenta and masses, can be define as
\begin{align}
J_{N}^{\left(  d\right)  }\left(  \left\{  k_{i}\right\}  ,\left\{
m_{i}\right\}  \right)   &  =\int\frac{d^{d}k}{(2\pi)^{d}}I_{N}\left(
k_{1},..,k_{N};m_{1},...,m_{N}\right)  ,\label{I1}\\
I_{N}\left(  k_{1},..,k_{N};m_{1},...,m_{N}\right)   &  =%
{\displaystyle\prod\limits_{i=1}^{N}}
\frac{1}{P\left(  k+k_{i},m_{i}\right)  },
\end{align}
where $P\left(  k+k_{i},m_{i}\right)  =\left(  k+k_{i}\right)  ^{2}-m_{i}^{2}%
$. The arbitrary internal momenta $k_{i}$ are related, in physical amplitudes,
to the external ones through their differences. At a specific spacetime
dimension, some of the structures defined above may present a divergent
character. Then, for their explicit evaluation, we have to specify some
prescription to deal with the mathematical objects which are not well-defined.
Usually the calculations become reliable only after adopting a regularization
technique. Such a procedure invariably modify the integrand in order to get a
convergent integral. Unfortunately, as is well known, the final results are,
in general, regularization dependent. This usually means that it is not
possible to specify, in a clear way, what are the particular effects of the
adopted regularization for the results or, in other words, to know precisely
in what extension the expression obtained is dependent in the used technique.
Beside that, there is an ambiguity relative to the kind of regularization
prescription which is chosen. Two different choices for the regularization can
lead to different results for the calculated integrals. These kinds of issues,
associated with regularizations prescriptions, are very well-known in the
corresponding literature.

On the other hand, constructed to be an alternative to the standard
regularization techniques, the method that we adopt in this paper aims to
avoid, as much as possible, specific choices in intermediary steps of
calculations, in such a way that all the possibilities still remain contained
in the final results. This goal can be accomplished by following a simple
sequence of steps. First, before introducing the integration sign, which can
be thought representing the last Feynman rule, we make a power counting of
loop momentum in order to get the superficial degree of divergence of the
integral, focusing on a particular spacetime dimension. After that, we can
rewrite the integrand $I_{N}\left(  k_{1},..,k_{N};m_{1},...,m_{N}\right)  $
by using an alternative representation for $P^{-1}\left(  k+k_{i}%
,m_{i}\right)  $, say%
\begin{align}
\frac{1}{P\left(  k+k_{i},m_{i}\right)  }  &  =\sum_{j=0}^{N}\frac{\left(
-1\right)  ^{j}\left(  k_{i}^{2}+2k_{i}\cdot k+\lambda^{2}-m_{i}^{2}\right)
^{j}}{\left(  k^{2}-\lambda^{2}\right)  ^{j+1}}\nonumber\\
&  +\frac{\left(  -1\right)  ^{N+1}\left(  k_{i}^{2}+2k_{i}\cdot k+\lambda
^{2}-m_{i}^{2}\right)  ^{N+1}}{\left(  k^{2}-\lambda^{2}\right)  ^{N+1}\left[
\left(  k+k_{i}\right)  ^{2}-m_{i}^{2}\right]  }, \label{identity}%
\end{align}
where the summation variable $N$ is taken as equal or major than the
superficial degree of divergence found. The arbitrary $\lambda$ parameter has
dimension of mass and plays the role of a mass scale. As would be expected,
the expression above is an identity and the expression on the right hand side
is really independent of the $\lambda$. After this reorganization of the
integrand, we can take the integration over the loop momentum $k$. As a
result, we have rewritten the original integral as a sum of finite as well as
divergent new integrals. The unique assumption is that the linearity in the
integration operation is a valid property for Feynman integrals. An important
point about this reorganization is that the internal momenta dependent parts
of the integrals are located only in finite ones. These finite integrals can
be evaluated without restrictions and the divergent ones are just rewritten in
terms of standard objects, conveniently defined. In this sense, the above
identity is just one among many others which could play the same role. The
procedure can be better visualized through a few examples, which we describe next.

Let us consider first the one-point integral in two dimensions which is given
by (from now on we will hide the arguments of $J_{N}^{\left(  d\right)  } $
always that they are not essential)%
\begin{equation}
J_{1}^{\left(  2\right)  }=\int\frac{d^{2}k}{(2\pi)^{4}}I_{1}\left(
k_{1},m_{1}\right)  \ , \label{J1-2}%
\end{equation}
which has a logarithmic degree of divergence. For this degree of divergence,
it is enough to take $N=0$ in Eq. (\ref{identity}), such that an alternative
representation for the integrand $I_{1}\left(  k_{1},m_{1}\right)  $ can be
written as
\begin{equation}
I_{1}\left(  k_{1},m_{1}\right)  =\frac{1}{\left[  P\left(  k,\lambda\right)
\right]  }-\frac{\left(  k_{1}^{2}+2k\cdot k_{1}+\lambda^{2}-m_{1}^{2}\right)
}{\left[  P\left(  k,\lambda\right)  \right]  \left[  P\left(  k+k_{1}%
,m_{1}\right)  \right]  },
\end{equation}
which, after substitution in (\ref{J1-2}), gives%
\begin{equation}
J_{1}^{\left(  2\right)  }=\int\frac{d^{2}k}{(2\pi)^{2}}\frac{1}{\left[
P\left(  k,\lambda\right)  \right]  }-\int\frac{d^{2}k}{(2\pi)^{2}}%
\frac{\left(  k_{1}^{2}+2k\cdot k_{1}+\lambda^{2}-m_{1}^{2}\right)  }{\left[
P\left(  k,\lambda\right)  \right]  \left[  P\left(  k+k_{1},m_{1}\right)
\right]  }.
\end{equation}
The last integral is finite and its integration can be easily done. We get%
\begin{equation}
J_{1}^{\left(  2\right)  }=\int\frac{d^{2}k}{(2\pi)^{2}}\frac{1}{\left[
P\left(  k,\lambda\right)  \right]  }-\frac{i}{\left(  4\pi\right)  }%
\ln\left(  \frac{m_{1}^{2}}{\lambda^{2}}\right)  \ .
\end{equation}
The first term, in the equation above, is a representative of a divergent
integral belonging to a class of Feynman integrals which we denominate basic
divergent integrals. Such class of integrals are characterized by the absence
of physical parameter in its integrands such that, in this sense, carry no
physical content. Following the main philosophy of the method, which means
does not perform the integration operation of these divergent structures, we
keep this object untouched and, for a better systematization of the results,
we define a $d$-dimensional basic logarithmically divergent object, given by%
\begin{equation}
I_{\log}^{\left(  d\right)  }\left(  \lambda^{2}\right)  =\int\frac{d^{d}%
k}{(2\pi)^{d}}\frac{1}{\left[  P\left(  k,\lambda\right)  \right]  ^{\frac
{d}{2}}}\ .
\end{equation}
Given this definition we can write
\begin{equation}
J_{1}^{\left(  2\right)  }=\left[  I_{\log}^{\left(  2\right)  }\left(
\lambda^{2}\right)  \right]  -\frac{i}{\left(  4\pi\right)  }\ln\left(
\frac{m_{1}^{2}}{\lambda^{2}}\right)  \ .
\end{equation}
The same integral, treated in four dimensions, namely $J_{1}^{\left(
4\right)  }$, has a quadratic degree of divergence. This require, at least, to
take $N=2$ in (\ref{identity}), which gives%
\begin{align}
I_{1}\left(  k_{1},m_{1}\right)   &  =\frac{1}{\left[  P\left(  k,\lambda
\right)  \right]  }-\frac{\left(  k_{1}^{2}+\lambda^{2}-m^{2}\right)
}{\left[  P\left(  k,\lambda\right)  \right]  ^{2}}+4k_{1}^{\alpha}%
k_{1}^{\beta}\frac{k_{\alpha}k_{\beta}}{\left[  P\left(  k,\lambda\right)
\right]  ^{3}}\nonumber\\
&  +\frac{\left(  k_{1}^{2}+\lambda^{2}-m^{2}\right)  ^{2}}{\left[  P\left(
k,\lambda\right)  \right]  ^{3}}-\frac{\left(  k_{1}^{2}+\lambda^{2}%
-m^{2}+2k\cdot k_{1}\right)  ^{3}}{\left[  P\left(  k,\lambda\right)  \right]
^{3}\left[  P\left(  k+k_{1},m_{1}\right)  \right]  }\ .
\end{align}
After introducing the integration over the loop momentum, we note that the
last two terms gives finite integrals. The final expression can be put in the
form%
\begin{align}
J_{1}^{\left(  4\right)  }  &  =k_{1}^{\alpha}k_{1}^{\beta}\left[
\Delta_{\alpha\beta}^{\left(  4\right)  }\left(  \lambda^{2}\right)  \right]
+\left[  I_{quad}^{\left(  4\right)  }\left(  \lambda^{2}\right)  \right]
+\left(  m_{1}^{2}-\lambda^{2}\right)  \left[  I_{\log}^{\left(  4\right)
}\left(  \lambda^{2}\right)  \right] \\
&  +\frac{i}{\left(  4\pi\right)  ^{2}}\left[  m_{1}^{2}-\lambda^{2}-m_{1}%
^{2}\ln\left(  \frac{m_{1}^{2}}{\lambda^{2}}\right)  \right]  \ ,\nonumber
\end{align}
where we have defined a $d$-dimensional basic quadratically divergent object,%
\begin{equation}
I_{quad}^{\left(  d\right)  }\left(  \lambda^{2}\right)  =\int\frac{d^{d}%
k}{\left(  2\pi\right)  ^{d}}\frac{1}{\left[  P\left(  k,\lambda\right)
\right]  ^{\frac{d}{2}-1}}\ ,
\end{equation}
and a tensorial object which can be viewed as a surface term because it can be
written as a total derivative%
\begin{align}
\Delta_{\mu\nu}^{\left(  d\right)  }\left(  \lambda^{2}\right)   &  =\int
\frac{d^{d}k}{\left(  2\pi\right)  ^{d}}\frac{\partial}{\partial k_{\mu}%
}\left(  -\frac{k_{\nu}}{\left[  P\left(  k,\lambda\right)  \right]
^{\frac{d}{2}}}\right)  \ ,\nonumber\\
&  =\int\frac{d^{d}k}{\left(  2\pi\right)  ^{d}}\left\{  d\frac{k_{\mu}k_{\nu
}}{\left[  P\left(  k,\lambda\right)  \right]  ^{\frac{d}{2}+1}}-\frac
{g_{\mu\nu}}{\left[  P\left(  k,\lambda\right)  \right]  ^{d}}\right\}  \ .
\end{align}
We should emphasize that neither of the above objects carry any physical
parameter. In six or more dimensions, where the divergences grow, the
procedure follows the same sequence of steps describe above, but require, if
one want to keep the described systematization, definitions of new basic
divergent objects similar to that showed above. In odd dimensions, the
procedure works in the same line, without additional hypothesis.

The scalar two-point function in two dimensions is finite but, in four
dimensions,%
\begin{equation}
J_{2}^{\left(  4\right)  }=\int\frac{d^{4}k}{(2\pi)^{4}}I_{2}\left(
k_{1},m_{1},k_{2},m_{2}\right)  \ ,
\end{equation}
it has a logarithmic divergence. Here we can apply identity (\ref{identity})
for each $P\left(  k+k_{i},m_{i}\right)  $ separately or for both
simultaneously. The last approach gives%
\begin{align}
I_{2}\left(  k_{1},m_{1},k_{2},m_{2}\right)   &  =\frac{1}{\left[  P\left(
k,\lambda\right)  \right]  ^{2}}-\frac{(k_{1}^{2}+2k_{1}\cdot k+\lambda
^{2}-m_{1}^{2})}{\left[  P\left(  k,\lambda\right)  \right]  ^{2}\left[
P\left(  k+k_{1},m_{1}\right)  \right]  }-\frac{(k_{2}^{2}+2k_{2}\cdot
k+\lambda^{2}-m_{1}^{2})}{\left[  P\left(  k,\lambda\right)  \right]
^{2}\left[  P\left(  k+k_{2},m_{2}\right)  \right]  }\nonumber\\
&  +\frac{(k_{1}^{2}+2k_{1}\cdot k+\lambda^{2}-m_{1}^{2})(k_{2}^{2}%
+2k_{2}\cdot k+\lambda^{2}-m_{2}^{2})}{\left[  P\left(  k,\lambda\right)
\right]  ^{2}\left[  P\left(  k+k_{1},m_{1}\right)  \right]  \left[  P\left(
k+k_{2},m_{2}\right)  \right]  }.
\end{align}
Only the first term will gives a divergent integral when the integration sign
is inserted. The remain ones are finite and, after their integration, we
obtain%
\begin{equation}
J_{2}^{\left(  4\right)  }=\left[  I_{\log}^{\left(  4\right)  }\left(
\lambda^{2}\right)  \right]  -\frac{i}{\left(  4\pi\right)  ^{2}}\left[
\xi_{0}^{\left(  0\right)  }\left(  m_{1}^{2};p^{2},m_{2}^{2};\lambda
^{2}\right)  \right]  \ ,
\end{equation}
with the finite part systematized by a set of functions defined in terms of an
integral over a Feynman parameter%
\begin{align}
\xi_{k}^{\left(  n\right)  }\left(  m_{1}^{2};p^{2},m_{2}^{2};\lambda
^{2}\right)   &  =\int_{0}^{1}dx\ x^{k}\left\{  \frac{\left[  Q\right]  ^{n}%
}{n!}\left[  \ln\left(  \frac{Q}{-\lambda^{2}}\right)  -\psi\left(
n+1\right)  -\gamma\right]  \right\}  ,\label{qsi}\\
Q\left(  m_{1}^{2};p^{2},m_{2}^{2},x\right)   &  =p^{2}x(1-x)+(m_{1}^{2}%
-m_{2}^{2})x-m_{1}^{2}\ ,\nonumber\\
\psi\left(  n+1\right)  +\gamma &  =%
{\displaystyle\sum\limits_{k=1}^{n}}
\frac{1}{k}\ ,\nonumber
\end{align}
with $\gamma$ being the Euler-Mascheroni constant and $n=0,1,2,\ldots$. The
integration over the Feynman parameter $x$ can be easily done and some
explicit expressions can be viewed in Ref. \cite{Orimar6}.

In its turn, this two-point integral in six dimensions has a quadratic degree
of divergence. Applying identity (\ref{identity}) for $N=2$ and integrating
the finite integrals allow us to write
\begin{align}
J_{2}^{\left(  6\right)  }  &  =-\frac{1}{3}\left[  \left(  k_{1}\right)
_{\alpha}\left(  k_{2}\right)  _{\beta}-2\left(  k_{2}+k_{1}\right)  _{\alpha
}\left(  k_{2}+k_{1}\right)  _{\beta}\right]  \left[  \Delta_{4;\alpha\beta
}^{\left(  6\right)  }\left(  \lambda^{2}\right)  \right] \nonumber\\
&  +\left[  I_{quad}^{\left(  6\right)  }\left(  \lambda^{2}\right)  \right]
-\frac{1}{6}\left[  p^{2}+3\left(  \lambda^{2}-m_{1}^{2}\right)  +3\left(
\lambda^{2}-m_{2}^{2}\right)  \right]  \left[  I_{\log}^{\left(  6\right)
}\left(  \lambda^{2}\right)  \right] \nonumber\\
&  +\frac{i}{\left(  4\pi\right)  ^{3}}\left[  \xi_{0}^{\left(  1\right)
}\left(  p^{2},m_{1}^{2},m_{2}^{2};\lambda^{2}\right)  \right]  \ .
\end{align}

The method's systematic is now clear. Given an arbitrary Feynman integral, we
can write an alternative representation of its integrand in such a way that,
when the sign integration is inserted, we find a sum of others Feynman
integrals. While the finite integrals can be performed and organized through a
set of well-defined functions conveniently defined (for one-loop divergent
integrals it is enough the set defined in Eq. (\ref{qsi})), the remaining
(divergent) part is composed by the following set of integrals
\begin{equation}
\int\frac{d^{d}k}{(2\pi)^{d}}\frac{\left\{  1;k_{\mu}k_{\nu};k_{\mu}k_{\nu
}k_{\alpha}k_{\beta};\ldots\right\}  }{\left[  P\left(  k,\lambda\right)
\right]  ^{N}}\ ,
\end{equation}
which are external momenta and masses independent. In this set, the tensor
integrals are reduced to a combination of scalar ones plus (tensorial) surface
terms. In the solution obtained, these integrals appear as coefficient of a
polynomial in the masses, external momenta and (arbitrary) internal routing
momenta $k_{i}$, as demonstrated in the cases treated above.

At this point one can ask: why such line of reasoning, for calculation and
manipulation of Feynman integrals, can be considered convenient as well as
useful? First of all, divergent mathematical structures are undefined
quantities, which are usually fixed by some choices necessarily made when a
regularization is adopted. In our approach, however, such indefiniteness are
still present at the final results through the basic divergent objects and
surface terms. In fact, the basic divergent objects do not need to be
evaluated since their coefficients, in a physical amplitude, are polynomials
in the external momenta such that they will invariably be absorbed in the
renormalization process. In a nonrenormalizable model, on the other hand, they
could be parameterized in order to fit the physical observables \cite{Orimar5}%
. In its turn, the coefficients of the surface terms are potentially ambiguous
if a generic choice for the labels of the internal lines momenta are made, as
one can see in the above expression obtained for $J_{2}^{\left(  6\right)  }$.
A series of investigations, made by the present authors and others, has
reveled that the relevant dependence of a perturbative calculation with a
regularization resides in the value attributed to these surface terms. In the
DR, for instance, these ambiguities are eliminated since such surface terms
are equal to zero. This is, in principle, an attractive possibility, but there
are others possible ones. The discussion of what are the consistency values
that should be attributed to such surface terms, if any exist, is long and is
out of the scope of the present paper. The point to be highlighted is that the
Feynman integrals, when evaluated through the method presented above, gives
results which preserve the arbitrariness involved because no choices are made
in the intermediary steps of the calculations. One obvious advantage of these
approach is that, when these results are used to calculate physical processes,
within a framework of a theory or model, such arbitrariness can be fixed
through choices guided by consistency requirements such as symmetries
maintenance and universality of calculations.

In the next section we will show how this approach can be used to calculate
multi-loop Feynman integrals by evaluating the two-loop sunrise integral in details.

\section{The equal-mass sunrise integral \label{Sec_SS}}

\begin{figure}[h]
\centering
\includegraphics[width=200pt,height=150pt]{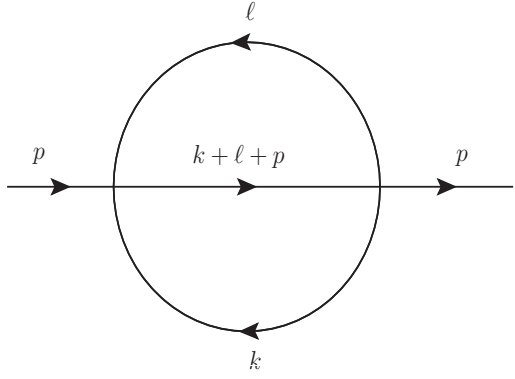} \caption{Sunrise
graph.}%
\label{fig_sunrise}%
\end{figure}

In this paper we are interested in the calculation of the well-known
$4$-dimensional equal mass sunrise diagram, which can be represented
schematically as in Fig. (\ref{fig_sunrise}).

It has an integral representation given by
\begin{align}
J_{SS}\left(  p,m\right)   &  =\int\frac{d^{4}k}{\left(  2\pi\right)  ^{4}%
}\frac{d^{4}\ell}{\left(  2\pi\right)  ^{4}}\left[  I_{SS}\left(
k,\ell,p,m\right)  \right]  \ ,\\
I_{SS}\left(  k,\ell,p,m\right)   &  =\frac{1}{\left[  P\left(  \ell,m\right)
\right]  \left[  P\left(  k,m\right)  \right]  \left[  P\left(  k+\ell
+p,m\right)  \right]  }\ .
\end{align}
Following the strategy discussed in the last section, the integrand $I_{SS}$
can be rewritten by applying recursively the identity (\ref{identity}), but
now written as
\begin{align}
\frac{1}{\left[  P\left(  k+\ell+p,m\right)  \right]  }  &  =\sum_{j=0}%
^{N}\frac{\left(  -1\right)  ^{j}\left[  p^{2}+2\left(  \ell+k\right)  \cdot
p\right]  ^{j}}{\left[  P\left(  k+\ell,m\right)  \right]  ^{j+1}}\nonumber\\
&  +\frac{\left(  -1\right)  ^{N+1}\left[  p^{2}+2\left(  \ell+k\right)  \cdot
p\right]  ^{N+1}}{\left[  P\left(  k+\ell,m\right)  \right]  ^{N+1}\left[
P\left(  k+\ell+p,m\right)  \right]  }. \label{identity2}%
\end{align}
From now on, we will hide the mass dependence in our notation, unless it
become necessary for the sake of clarity. Since the total degree of divergence
of $J_{SS}$ is quadratic, it is enough to choose $N=2$ in the above identity.
Thus%
\begin{align}
I_{SS}  &  =\frac{1}{\left[  P\left(  k\right)  \right]  \left[  P\left(
\ell\right)  \right]  \left[  P\left(  k+\ell\right)  \right]  }-p^{2}\frac
{1}{\left[  P\left(  k\right)  \right]  \left[  P\left(  \ell\right)  \right]
\left[  P\left(  k+\ell\right)  \right]  ^{2}}\nonumber\\
&  +4p^{\alpha}p^{\beta}\frac{\left(  \ell+k\right)  _{\alpha}\left(
\ell+k\right)  _{\beta}}{\left[  P\left(  k\right)  \right]  \left[  P\left(
\ell\right)  \right]  \left[  P\left(  k+\ell\right)  \right]  ^{3}}%
+\frac{p^{4}}{\left[  P\left(  k\right)  \right]  \left[  P\left(
\ell\right)  \right]  \left[  P\left(  k+\ell\right)  \right]  ^{3}%
}\nonumber\\
&  -\frac{\left[  p^{2}+2\left(  \ell+k\right)  \cdot p\right]  ^{3}}{\left[
P\left(  k\right)  \right]  \left[  P\left(  \ell\right)  \right]  \left[
P\left(  k+\ell\right)  \right]  ^{3}\left[  P\left(  k+\ell+p\right)
\right]  }\ ,
\end{align}
where we have discarded odd terms. By power counting, the last two terms above
are convergent, but still hide a subdivergence. If one integrate them over the
loop momenta, a divergence will emerge in the Feynman parameters integrals.
This a characteristic associated to certain types of two-loop Feynman
integrals. A careful analysis, however, reveals that there is a mutual
cancellation of such subdivergences coming from both referred terms. In order
to see that, it is convenient to first perform a shift ($\ell^{\prime}=\ell
+k$) and after use an identity similar to (\ref{identity}), i.e.,%
\begin{equation}
\frac{1}{\left[  P\left(  k+\ell\right)  \right]  }=\frac{1}{\left[  P\left(
k\right)  \right]  }-\frac{\left[  \ell^{2}-2\left(  \ell\cdot k\right)
\right]  }{\left[  P\left(  k\right)  \right]  \left[  P\left(  k+\ell\right)
\right]  }\ ,
\end{equation}
to write%
\begin{align}
J_{SS}  &  =\int\frac{d^{4}k}{\left(  2\pi\right)  ^{4}}\frac{d^{4}\ell
}{\left(  2\pi\right)  ^{4}}\frac{1}{\left[  P\left(  k\right)  \right]
\left[  P\left(  \ell\right)  \right]  \left[  P\left(  k+\ell\right)
\right]  }\nonumber\\
&  +p^{\alpha}p^{\beta}\int\frac{d^{4}k}{\left(  2\pi\right)  ^{4}}\frac
{d^{4}\ell}{\left(  2\pi\right)  ^{4}}\left\{  \frac{4\left(  \ell+k\right)
_{\alpha}\left(  \ell+k\right)  _{\beta}}{\left[  P\left(  k\right)  \right]
\left[  P\left(  \ell\right)  \right]  \left[  P\left(  k+\ell\right)
\right]  ^{3}}-\frac{g_{\alpha\beta}}{\left[  P\left(  k\right)  \right]
\left[  P\left(  \ell\right)  \right]  \left[  P\left(  k+\ell\right)
\right]  ^{2}}\right\} \nonumber\\
&  +\int\frac{d^{4}k}{\left(  2\pi\right)  ^{4}}\frac{1}{\left[  P\left(
k\right)  \right]  ^{2}}\frac{d^{4}\ell}{\left(  2\pi\right)  ^{4}}\left\{
\frac{p^{4}}{\left[  P\left(  \ell\right)  \right]  ^{3}}-\frac{\left[
p^{2}+2\ell\cdot p\right]  ^{3}}{\left[  P\left(  \ell\right)  \right]
^{3}\left[  P\left(  \ell+p\right)  \right]  }\right\} \nonumber\\
&  -\int\frac{d^{4}k}{\left(  2\pi\right)  ^{4}}\frac{d^{4}\ell}{\left(
2\pi\right)  ^{4}}\left\{  \frac{p^{4}}{\left[  P\left(  \ell\right)  \right]
^{3}}-\frac{\left[  p^{2}+2\ell\cdot p\right]  ^{3}}{\left[  P\left(
\ell\right)  \right]  ^{3}\left[  P\left(  \ell+p\right)  \right]  }\right\}
\frac{\left[  \ell^{2}-2\left(  \ell\cdot k\right)  \right]  }{\left[
P\left(  k\right)  \right]  ^{2}\left[  P\left(  \ell-k\right)  \right]  }\ .
\label{Iss1}%
\end{align}
But, it is straightforward to see that
\begin{equation}
\int\frac{d^{4}\ell}{\left(  2\pi\right)  ^{4}}\left\{  \frac{p^{4}}{\left[
P\left(  \ell\right)  \right]  ^{3}}-\frac{\left[  p^{2}+2\ell\cdot p\right]
^{3}}{\left[  P\left(  \ell\right)  \right]  ^{3}\left[  P\left(
\ell+p\right)  \right]  }\right\}  =0\ ,
\end{equation}
such that the referred cancellation occurs.

The integration over the loop momenta in the last term in Eq. (\ref{Iss1}) can
easily be done and the result can be written as%
\begin{align}
J_{SS}  &  =\int\frac{d^{4}k}{\left(  2\pi\right)  ^{4}}\frac{d^{4}\ell
}{\left(  2\pi\right)  ^{4}}\frac{1}{\left[  P\left(  k\right)  \right]
\left[  P\left(  \ell\right)  \right]  \left[  P\left(  k+\ell\right)
\right]  }\nonumber\\
&  +p^{\alpha}p^{\beta}\int\frac{d^{4}k}{\left(  2\pi\right)  ^{4}}\frac
{d^{4}\ell}{\left(  2\pi\right)  ^{4}}\left\{  \frac{4\left(  \ell+k\right)
_{\alpha}\left(  \ell+k\right)  _{\beta}}{\left[  P\left(  k\right)  \right]
\left[  P\left(  \ell\right)  \right]  \left[  P\left(  k+\ell\right)
\right]  ^{3}}-\frac{g_{\alpha\beta}}{\left[  P\left(  k\right)  \right]
\left[  P\left(  \ell\right)  \right]  \left[  P\left(  k+\ell\right)
\right]  ^{2}}\right\} \nonumber\\
&  -\frac{1}{\left(  4\pi\right)  ^{4}}\frac{2p^{4}}{m^{2}}\int_{0}%
^{1}dx\ \frac{x\left(  1-2x\right)  }{\left[  x\left(  1-x\right)  -1\right]
}\left\{  \frac{1}{2}-\frac{1}{\left[  x\left(  1-x\right)  -1\right]  }%
+\frac{\ln\left[  x\left(  1-x\right)  \right]  }{\left[  x\left(  1-x\right)
-1\right]  ^{2}}\right\} \nonumber\\
&  -\frac{p^{2}}{\left(  4\pi\right)  ^{4}}\int_{0}^{1}dx\int_{0}^{1}%
dy\ \frac{\left(  1-2x\right)  \left(  1-2y\right)  }{\left(  1-x\right)
\left[  x\left(  1-x\right)  -1\right]  }\ln\left[  \frac{p^{2}y\left(
1-y\right)  +\left(  m^{2}-\mu^{2}\right)  y-m^{2}}{\left(  m^{2}-\mu
^{2}\right)  y-m^{2}}\right]  \ , \label{Jss1}%
\end{align}
with the definition $\mu^{2}=\frac{m^{2}}{x\left(  1-x\right)  }$. In the
multi-loop calculations, the integration over the Feynman parameters, in a
closed form, may represents a new challenge to be overcome. In one-loop
calculations, it is well-known that the results can be written in terms of
MPLs \cite{Goncharov, Goncharov2, Remiddi4, Vollinga}. On the other hand, for
integrals beyond one-loop, this is not more possible for all of them. In our
case, the integrals over the Feynman parameters showed above can be performed
both numerically or analytically by using eMPLs \cite{Beilinson, Levin, Brown,
Broedel, Broedel2, Broedel3, Broedel4}. We will done this task in a separate
section ahead.

As a main feature of the method, the remaining divergences (the two first
lines in the equation above) have been isolated into integrals that do not
contain physical momenta $p$ inside them, for such a reason called basic
divergent ones. The physical mass $m$, at this stage of calculation, still
inside of these integrands, can be replaced by an arbitrary mass scale
$\lambda$ by using the so called scale relations \cite{Orimar3}, which will be
discussed in the Section (\ref{Sec_Scale_Relations}). Such objects are not
well-defined and could not be integrated out without a regularization or
similar prescription. In the one-loop integrals, we shown that such terms can
be reduced to scalar objects plus tensorial surface terms. As we will show, in
the case of two-loop sunrise graph the same can be done.

The prescription applied above to calculate $J_{SS}$ is exactly the same one
used to calculate the one-loop integrals, in the Section (\ref{sec_method}).
However, the method is flexible enough to allows us to use different
approaches too. In order to see that, let us recover the result (\ref{Jss1})
but now following another possible path to deal with the $J_{SS}$ integral
within the method. As we will show soon, both results will be equivalent, but
having a slightly different representation for the finite integrals, which
will be an advantage when the Feynman parameters integrals are performed in
the Section (\ref{sec_IOFP}). In this second approach, before applying the
identity (\ref{identity2}), we first rewrite the integrand by using some
identities constructed through the method of integration-by-parts, which is a
standard tool used in Feynman loop calculations. First let us consider two
trivial total derivatives%
\begin{align}
\frac{\partial}{\partial k_{\mu}}\left\{  \frac{k_{\mu}}{\left[  P\left(
k\right)  \right]  \left[  P\left(  \ell\right)  \right]  \left[  P\left(
k+\ell+p\right)  \right]  }\right\}   &  =\frac{4}{\left[  P\left(  k\right)
\right]  \left[  P\left(  \ell\right)  \right]  \left[  P\left(
k+\ell+p\right)  \right]  }\nonumber\\
&  -\frac{2k^{2}}{\left[  P\left(  k\right)  \right]  ^{2}\left[  P\left(
\ell\right)  \right]  \left[  P\left(  k+\ell+p\right)  \right]  }\nonumber\\
&  -\frac{2k\cdot\left(  k+\ell+p\right)  }{\left[  P\left(  k\right)
\right]  \left[  P\left(  \ell\right)  \right]  \left[  P\left(
k+\ell+p\right)  \right]  ^{2}}\ , \label{TD1}%
\end{align}%
\begin{align}
\frac{\partial}{\partial\ell_{\mu}}\left\{  \frac{\ell_{\mu}}{\left[  P\left(
k\right)  \right]  \left[  P\left(  \ell\right)  \right]  \left[  P\left(
k+\ell+p\right)  \right]  }\right\}   &  =\frac{4}{\left[  P\left(  k\right)
\right]  \left[  P\left(  \ell\right)  \right]  \left[  P\left(
k+\ell+p\right)  \right]  }\nonumber\\
&  -\frac{2\ell^{2}}{\left[  P\left(  k\right)  \right]  \left[  P\left(
\ell\right)  \right]  ^{2}\left[  P\left(  k+\ell+p\right)  \right]
}\nonumber\\
&  -\frac{2\ell\cdot\left(  k+\ell+p\right)  }{\left[  P\left(  k\right)
\right]  \left[  P\left(  \ell\right)  \right]  \left[  P\left(
k+\ell+p\right)  \right]  ^{2}}\ . \label{TD2}%
\end{align}
Summing up the above expressions and reducing the bilinears in the numerator
gives%
\begin{align}
I_{SS}  &  =\Delta I_{SS}+\frac{3m^{2}}{\left[  P\left(  k\right)  \right]
\left[  P\left(  \ell\right)  \right]  \left[  P\left(  k+\ell+p\right)
\right]  ^{2}}\nonumber\\
&  -\frac{p\cdot\left(  k+\ell+p\right)  }{\left[  P\left(  k\right)  \right]
\left[  P\left(  \ell\right)  \right]  \left[  P\left(  k+\ell+p\right)
\right]  ^{2}}\ ,
\end{align}
with the definition%
\begin{align}
\Delta I_{SS}  &  =\frac{1}{2}\frac{\partial}{\partial k_{\mu}}\left\{
\frac{k_{\mu}}{\left[  P\left(  k\right)  \right]  \left[  P\left(
\ell\right)  \right]  \left[  P\left(  k+\ell+p\right)  \right]  }\right\}
\nonumber\\
&  +\frac{1}{2}\frac{\partial}{\partial\ell_{\mu}}\left\{  \frac{\ell_{\mu}%
}{\left[  P\left(  k\right)  \right]  \left[  P\left(  \ell\right)  \right]
\left[  P\left(  k+\ell+p\right)  \right]  }\right\}  \ .
\end{align}
Through this step, we have reduced the integrand into surface terms plus two
others terms which have logarithmic and linear degree of divergence,
respectively. Next, let us apply the identity (\ref{identity2}) in order to
rewrite these two terms, in the same spirit of the discussion above. We get%
\begin{align}
\frac{1}{\left[  P\left(  k\right)  \right]  \left[  P\left(  \ell\right)
\right]  \left[  P\left(  k+\ell+p\right)  \right]  ^{2}}  &  =\frac
{1}{\left[  P\left(  k\right)  \right]  \left[  P\left(  \ell\right)  \right]
\left[  P\left(  k+\ell\right)  \right]  ^{2}}\nonumber\\
&  -\frac{\left[  p^{2}+2\left(  \ell+k\right)  \cdot p\right]  }{\left[
P\left(  k\right)  \right]  ^{2}\left[  P\left(  \ell\right)  \right]  \left[
P\left(  k+\ell\right)  \right]  \left[  P\left(  k+\ell+p\right)  \right]
}\ ,
\end{align}%
\begin{align}
\frac{p\cdot\left(  k+\ell+p\right)  }{\left[  P\left(  k\right)  \right]
\left[  P\left(  \ell\right)  \right]  \left[  P\left(  k+\ell+p\right)
\right]  ^{2}}  &  =\frac{2\left(  p\cdot k\right)  \left[  p\cdot\left(
\ell+k\right)  \right]  }{\left[  P\left(  \ell\right)  \right]  \left[
P\left(  k\right)  \right]  ^{2}\left[  P\left(  k+\ell\right)  \right]  ^{2}%
}\nonumber\\
&  -\frac{\left(  p\cdot k\right)  \left[  p^{2}+2\left(  \ell+k\right)  \cdot
p\right]  ^{2}}{\left[  P\left(  \ell\right)  \right]  \left[  P\left(
k\right)  \right]  ^{2}\left[  P\left(  k+\ell\right)  \right]  ^{2}\left[
P\left(  k+\ell+p\right)  \right]  }\ .
\end{align}
Then%
\begin{align}
J_{SS}  &  =\Delta J_{SS}+3m^{2}\int\frac{d^{4}k}{\left(  2\pi\right)  ^{4}%
}\frac{d^{4}\ell}{\left(  2\pi\right)  ^{4}}\frac{1}{\left[  P\left(
k\right)  \right]  \left[  P\left(  \ell\right)  \right]  \left[  P\left(
k+\ell\right)  \right]  ^{2}}\nonumber\\
&  -2\int\frac{d^{4}k}{\left(  2\pi\right)  ^{4}}\frac{d^{4}\ell}{\left(
2\pi\right)  ^{4}}\frac{\left(  p\cdot k\right)  \left[  p\cdot\left(
\ell+k\right)  \right]  }{\left[  P\left(  \ell\right)  \right]  \left[
P\left(  k\right)  \right]  ^{2}\left[  P\left(  k+\ell\right)  \right]  ^{2}%
}\nonumber\\
&  -3m^{2}\int\frac{d^{4}\ell}{\left(  2\pi\right)  ^{4}}\frac{\left(
p^{2}-2\ell\cdot p\right)  }{\left[  P\left(  \ell\right)  \right]  \left[
P\left(  \ell-p\right)  \right]  }\int\frac{d^{4}k}{\left(  2\pi\right)  ^{4}%
}\frac{1}{\left[  P\left(  k\right)  \right]  ^{2}\left[  P\left(
k+\ell\right)  \right]  }\nonumber\\
&  +p^{\mu}\int\frac{d^{4}\ell}{\left(  2\pi\right)  ^{4}}\frac{\left[
p^{2}+2\left(  \ell\cdot p\right)  \right]  ^{2}}{\left[  P\left(
\ell\right)  \right]  ^{2}\left[  P\left(  \ell+p\right)  \right]  }\int
\frac{d^{4}k}{\left(  2\pi\right)  ^{4}}\frac{k_{\mu}}{\left[  P\left(
k\right)  \right]  ^{2}\left[  P\left(  \ell-k\right)  \right]  }\ ,
\end{align}
where $\Delta J_{SS}=\int\frac{d^{4}k}{\left(  2\pi\right)  ^{4}}\frac
{d^{4}\ell}{\left(  2\pi\right)  ^{4}}\Delta I_{SS}\ $. Now, the last two
integrals are convergent and can be easily done.

Let us work out the surface term $\Delta J_{SS}$ by using again the identity
(\ref{identity2}). After this, we see that it is, in fact, independent of the
momentum $p$, i.e.,
\begin{align}
\Delta J_{SS}  &  =\frac{1}{2}\int\frac{d^{4}k}{\left(  2\pi\right)  ^{4}%
}\frac{d^{4}\ell}{\left(  2\pi\right)  ^{4}}\frac{\partial}{\partial k_{\mu}%
}\left\{  \frac{k_{\mu}}{\left[  P\left(  k\right)  \right]  \left[  P\left(
\ell\right)  \right]  \left[  P\left(  k+\ell\right)  \right]  }\right\}
\nonumber\\
&  +\frac{1}{2}\int\frac{d^{4}k}{\left(  2\pi\right)  ^{4}}\frac{d^{4}\ell
}{\left(  2\pi\right)  ^{4}}\frac{\partial}{\partial\ell_{\mu}}\left\{
\frac{\ell_{\mu}}{\left[  P\left(  k\right)  \right]  \left[  P\left(
\ell\right)  \right]  \left[  P\left(  k+\ell\right)  \right]  }\right\}  \ .
\end{align}
By taking the derivatives and eliminating the bilinears which appears in the
numerator we get%
\begin{equation}
\Delta J_{SS}=\int\frac{d^{4}\ell}{\left(  2\pi\right)  ^{4}}\frac{d^{4}%
k}{\left(  2\pi\right)  ^{4}}\left\{  \frac{1}{\left[  P\left(  k\right)
\right]  \left[  P\left(  \ell\right)  \right]  \left[  P\left(
k+\ell\right)  \right]  }-\frac{3m^{2}}{\left[  P\left(  k\right)  \right]
\left[  P\left(  \ell\right)  \right]  \left[  P\left(  k+\ell\right)
\right]  ^{2}}\right\}  \ ,
\end{equation}
and%
\begin{align}
J_{SS}  &  =\int\frac{d^{4}\ell}{\left(  2\pi\right)  ^{4}}\frac{d^{4}%
k}{\left(  2\pi\right)  ^{4}}\frac{1}{\left[  P\left(  k\right)  \right]
\left[  P\left(  \ell\right)  \right]  \left[  P\left(  k+\ell\right)
\right]  }\nonumber\\
&  -2p^{\mu}p^{\nu}\int\frac{d^{4}k}{\left(  2\pi\right)  ^{4}}\frac{d^{4}%
\ell}{\left(  2\pi\right)  ^{4}}\frac{k_{\mu}\left(  \ell+k\right)  _{\nu}%
}{\left[  P\left(  \ell\right)  \right]  \left[  P\left(  k\right)  \right]
^{2}\left[  P\left(  k+\ell\right)  \right]  ^{2}}\nonumber\\
&  +\frac{1}{\left(  4\pi\right)  ^{4}}\int_{0}^{1}dy\int_{0}^{1}dx\ \left[
\frac{3m^{2}}{x}-p^{2}\left(  1-y\right)  \right]  \ \ln\left[  \frac{{p}%
^{2}y\left(  1-y\right)  +\left(  m^{2}-\mu^{2}\right)  y-m^{2}}{\left(
m^{2}-\mu^{2}\right)  y-m^{2}}\right]  \ .
\end{align}
Finally, by noting the obvious identity
\begin{align}
&  \int\frac{d^{4}k}{\left(  2\pi\right)  ^{4}}\frac{d^{4}\ell}{\left(
2\pi\right)  ^{4}}\frac{2k_{\mu}\left(  \ell+k\right)  _{\nu}}{\left[
P\left(  \ell\right)  \right]  \left[  P\left(  k\right)  \right]  ^{2}\left[
P\left(  k+\ell\right)  \right]  ^{2}}\nonumber\\
&  =-\int\frac{d^{4}k}{\left(  2\pi\right)  ^{4}}\frac{d^{4}\ell}{\left(
2\pi\right)  ^{4}}\left\{  \frac{4\left(  \ell+k\right)  _{\mu}\left(
\ell+k\right)  _{\nu}}{\left[  P\left(  k\right)  \right]  \left[  P\left(
\ell\right)  \right]  \left[  P\left(  k+\ell\right)  \right]  ^{3}}%
-\frac{g_{\mu\nu}}{\left[  P\left(  k\right)  \right]  \left[  P\left(
\ell\right)  \right]  \left[  P\left(  k+\ell\right)  \right]  ^{2}}\right\}
\ ,
\end{align}
one can see that the above formula for $J_{SS}$ is, in fact, similar to the
Eq. (\ref{Jss1}), but with the finite part written in a slightly different
form. In this form, the task of write the result of $J_{SS}$ in terms of
eMPLs, in the Sec. (\ref{sec_IOFP}), will become simpler.

Performing the (finite) integration over momentum $\ell$, in the integral on
the right hand side of the above equation, allow us to find that its divergent
part can be written as one-loop basic objects, i.e.,%
\begin{align}
&  \int\frac{d^{4}k}{\left(  2\pi\right)  ^{4}}\frac{d^{4}\ell}{\left(
2\pi\right)  ^{4}}\frac{2k_{\mu}\left(  \ell+k\right)  _{\nu}}{\left[
P\left(  \ell\right)  \right]  \left[  P\left(  k\right)  \right]  ^{2}\left[
P\left(  k+\ell\right)  \right]  ^{2}}\nonumber\\
&  =\frac{i}{32\pi^{2}}\left\{  g_{\mu\nu}\left[  I_{\log}^{\left(  4\right)
}\left(  m^{2}\right)  \right]  +\left[  \Delta_{\mu\nu}^{\left(  4\right)
}\left(  m^{2}\right)  \right]  \right\} \nonumber\\
&  -\frac{1}{\left(  4\pi\right)  ^{4}}\frac{g_{\mu\nu}}{4}\int_{0}%
^{1}dx\ \left\{  1-\frac{2}{\left[  x\left(  1-x\right)  -1\right]  }%
+\frac{2\ln\left[  x\left(  1-x\right)  \right]  }{\left[  x\left(
1-x\right)  -1\right]  ^{2}}\right\}  \ ,
\end{align}
such that%
\begin{align}
J_{SS}  &  =\left[  I_{quad}^{\left(  2L\right)  }\left(  m^{2}\right)
\right]  -\frac{i}{32\pi^{2}}\left\{  p^{2}\left[  I_{\log}^{\left(  4\right)
}\left(  m^{2}\right)  \right]  +p^{\mu}p^{\nu}\left[  \Delta_{\mu\nu
}^{\left(  4\right)  }\left(  m^{2}\right)  \right]  \right\} \nonumber\\
&  +\frac{1}{\left(  4\pi\right)  ^{4}}\frac{p^{2}}{4}\int_{0}^{1}dx\ \left\{
1-\frac{2}{\left[  x\left(  1-x\right)  -1\right]  }+\frac{2\ln\left[
x\left(  1-x\right)  \right]  }{\left[  x\left(  1-x\right)  -1\right]  ^{2}%
}\right\} \nonumber\\
&  +\frac{1}{\left(  4\pi\right)  ^{4}}\int_{0}^{1}dy\int_{0}^{1}dx\ \left[
\frac{3m^{2}}{x}-p^{2}\left(  1-y\right)  \right]  \ \ln\left[  \frac{{p}%
^{2}y\left(  1-y\right)  +\left(  m^{2}-\mu^{2}\right)  y-m^{2}}{\left(
m^{2}-\mu^{2}\right)  y-m^{2}}\right]  \ , \label{Jss2}%
\end{align}
where we defined a typical $4$-dimensional two-loop basic object with
quadratic degree of divergence
\begin{equation}
I_{quad}^{\left(  2L\right)  }\left(  m^{2}\right)  =\int\frac{d^{4}k}{\left(
2\pi\right)  ^{4}}\frac{d^{4}\ell}{\left(  2\pi\right)  ^{4}}\frac{1}{\left[
P\left(  k\right)  \right]  \left[  P\left(  \ell\right)  \right]  \left[
P\left(  k+\ell\right)  \right]  }\ .
\end{equation}
In the previous formula for $J_{SS}$, one can see that the undefined
quantities are coefficient of a polynomial in momentum $p$, which allows us to
absorb them in a renormalization procedure, if this sunrise integral would
part of a physical process prediction made within a renormalizable theory. In
this sense, such undefined objects do not require any explicit calculation
and, then, for practical purposes no regularization is required.

In order to complete the calculation of $J_{SS}$, we need to perform the
integration over the Feynman parameters and also its numerical analysis. We
will perform this task in the section (\ref{sec_IOFP}). Before that, let us
discuss the scale properties of the two-loop basic divergent objects.

\section{Scale relations for two-loops scalar basic divergent objects
\label{Sec_Scale_Relations}}

In the last section, we have defined the basic divergent objects using the
physical mass $m$. However, the mass parameter used to define the basic
divergent objects is arbitrary because it can be chosen freely in the
separation process made through the identity (\ref{identity2}). Thus, it is
possible to find relations connecting such objects defined at different mass
scales, which are called scale relations. For objects typical of one-loop,
such relations are well-known and they are discussed in Refs. \cite{Orimar3,
Orimar5}. In this section we will show the scale relations for two typical
two-loop scalar divergent objects, having logarithm and quadratic degree of
divergence, respectively.

\subsection{Logarithmic divergence}

Let us start with the logarithmic one defined by%
\begin{equation}
I_{\log}^{\left(  2L\right)  }\left(  \lambda^{2}\right)  =\int\frac{d^{4}%
k}{\left(  2\pi\right)  ^{4}}\int\frac{d^{4}\ell}{\left(  2\pi\right)  ^{4}%
}\frac{1}{\left[  P\left(  k,\lambda\right)  \right]  \left[  P\left(
\ell,\lambda\right)  \right]  \left[  P\left(  \ell+k,\lambda\right)  \right]
^{2}}\ ,
\end{equation}
where $\lambda$ is an arbitrary mass which may play a role of a mass scale. It
does not appears in the $J_{SS}$ result, but it will be present in other
two-loops Feynman diagrams. Differentiation with respect to $\lambda^{2}$
gives
\begin{align}
\frac{\partial}{\partial\lambda^{2}}\left[  I_{\log}^{\left(  2L\right)
}\left(  \lambda^{2}\right)  \right]   &  =\int\frac{d^{4}k}{\left(
2\pi\right)  ^{4}}\int\frac{d^{4}\ell}{\left(  2\pi\right)  ^{4}}\frac
{2}{\left(  k^{2}-\lambda^{2}\right)  ^{2}\left(  \ell^{2}-\lambda^{2}\right)
\left[  \left(  \ell+k\right)  ^{2}-\lambda^{2}\right]  ^{2}}\nonumber\\
&  +\int\frac{d^{4}k}{\left(  2\pi\right)  ^{4}}\int\frac{d^{4}\ell}{\left(
2\pi\right)  ^{4}}\frac{2}{\left(  k^{2}-\lambda^{2}\right)  \left(  \ell
^{2}-\lambda^{2}\right)  \left[  \left(  \ell+k\right)  ^{2}-\lambda
^{2}\right]  ^{3}}\ .
\end{align}
The last integral is finite by power counting but still contains a
subdivergence. To see that, we first perform a shift ($\ell+k=k^{\prime}$) and
apply the identity%
\begin{equation}
\frac{1}{\left[  \left(  k-{\ell}\right)  ^{2}-\lambda^{2}\right]  }=\frac
{1}{\left(  \ell^{2}-\lambda^{2}\right)  }-\frac{\left[  {k}^{2}-2\left(
k\cdot{\ell}\right)  \right]  }{\left(  \ell^{2}-\lambda^{2}\right)  \left[
\left(  k-{\ell}\right)  ^{2}-\lambda^{2}\right]  }\ ,
\end{equation}
to get%
\begin{equation}
\frac{\partial}{\partial\lambda^{2}}\left[  I_{\log}^{\left(  2L\right)
}\left(  \lambda^{2}\right)  \right]  =-\frac{i}{\left(  4\pi\right)  ^{2}%
}\frac{{1}}{\lambda^{2}}\left[  I_{\log}^{\left(  4\right)  }\left(
\lambda^{2}\right)  \right]  +\frac{1}{\left(  4\pi\right)  ^{4}}\frac
{1}{\lambda^{2}}\ .
\end{equation}
Using the scale relation for (one-loop) $I_{\log}^{\left(  4\right)  }\left(
\lambda^{2}\right)  $ (see Ref. \cite{Orimar3}), i.e.,
\begin{equation}
I_{\log}^{\left(  4\right)  }\left(  \lambda^{2}\right)  =I_{\log}^{\left(
4\right)  }\left(  \lambda_{0}^{2}\right)  +\frac{i}{\left(  4\pi\right)
^{2}}\ln\left(  \frac{\lambda_{0}^{2}}{\lambda^{2}}\right)  \ ,
\end{equation}
where $\lambda_{0}$ is another arbitrary mass scale, and integrating on both
sides gives%
\begin{equation}
I_{\log}^{\left(  2L\right)  }\left(  \lambda^{2}\right)  =-\frac{i}{\left(
4\pi\right)  ^{2}}\left[  I_{\log}^{\left(  4\right)  }\left(  \lambda_{0}%
^{2}\right)  \right]  \ln\lambda^{2}+\frac{1}{\left(  4\pi\right)  ^{4}%
}\left[  \ln\left(  \lambda_{0}^{2}\right)  \ln\lambda^{2}-\frac{1}{2}\ln
^{2}\left(  \lambda^{2}\right)  -\ln\lambda^{2}\right]  +C_{1}\ ,
\end{equation}
with $C_{1}$ being a constant independent of the $\lambda$. By choosing
$\lambda=\lambda_{0}$ in the above expression%
\begin{equation}
I_{\log}^{\left(  2L\right)  }\left(  \lambda_{0}^{2}\right)  =-\frac
{i}{\left(  4\pi\right)  ^{2}}\left[  I_{\log}^{\left(  4\right)  }\left(
\lambda_{0}^{2}\right)  \right]  \ln\lambda_{0}^{2}+\frac{1}{\left(
4\pi\right)  ^{4}}\left[  \ln\left(  \lambda_{0}^{2}\right)  \ln\lambda
_{0}^{2}-\frac{1}{2}\ln^{2}\left(  \lambda_{0}^{2}\right)  -\ln\lambda_{0}%
^{2}\right]  +C_{1}\ ,
\end{equation}
and subtracting the last two equations gives%
\begin{equation}
I_{\log}^{\left(  2L\right)  }\left(  \lambda^{2}\right)  =I_{\log}^{\left(
2L\right)  }\left(  \lambda_{0}^{2}\right)  -\frac{i}{\left(  4\pi\right)
^{2}}\left\{  I_{\log}^{\left(  4\right)  }\left(  \lambda_{0}^{2}\right)
-\frac{i}{\left(  4\pi\right)  ^{2}}\left[  1+\frac{1}{2}\ln\left(
\frac{\lambda^{2}}{\lambda_{0}^{2}}\right)  \right]  \right\}  \ln\left(
\frac{\lambda^{2}}{\lambda_{0}^{2}}\right)  \ , \label{I_log_2L}%
\end{equation}
which is the scale relation searched for the object $I_{\log}^{\left(
2L\right)  }\left(  \lambda^{2}\right)  $.

\subsection{Quadratic divergence}

Next let us consider the object%
\begin{equation}
I_{quad}^{\left(  2L\right)  }\left(  \lambda^{2}\right)  =\int\frac{d^{4}%
k}{\left(  2\pi\right)  ^{4}}\int\frac{d^{4}\ell}{\left(  2\pi\right)  ^{4}%
}\frac{1}{\left(  k^{2}-\lambda^{2}\right)  \left(  \ell^{2}-\lambda
^{2}\right)  \left[  \left(  \ell+k\right)  ^{2}-\lambda^{2}\right]  }\ ,
\end{equation}
which has emerged in (\ref{Jss2}). By following the same previously strategy,
we first differentiate the above equation with respect to $\lambda^{2}$%
\begin{equation}
\frac{\partial}{\partial\lambda^{2}}\left[  I_{quad}^{\left(  2L\right)
}\left(  \lambda^{2}\right)  \right]  =3\left[  I_{\log}^{\left(  2L\right)
}\left(  \lambda^{2}\right)  \right]  \ ,
\end{equation}
use the scale relation obtained for $I_{\log}^{\left(  2L\right)  }\left(
\lambda^{2}\right)  $, Eq. (\ref{I_log_2L}), and after integrate on both sides
to get%
\begin{align}
I_{quad}^{\left(  2L\right)  }\left(  \lambda^{2}\right)   &  =3\lambda
^{2}\left[  I_{\log}^{\left(  2L\right)  }\left(  \lambda_{0}^{2}\right)
\right]  +\frac{i}{\left(  4\pi\right)  ^{2}}3\lambda^{2}\left[  I_{\log
}^{\left(  4\right)  }\left(  \lambda_{0}^{2}\right)  \right]  \left[
1-\ln\left(  \frac{\lambda^{2}}{\lambda_{0}^{2}}\right)  \right] \nonumber\\
&  -\frac{1}{\left(  4\pi\right)  ^{4}}\frac{3}{2}\lambda^{2}\ln^{2}\left(
\frac{\lambda^{2}}{\lambda_{0}^{2}}\right)  +C_{2}\ ,
\end{align}
where $C_{2}$ is a constant independent of $\lambda$. Taking $\lambda
=\lambda_{0}$ in the above formula gives%
\begin{equation}
I_{quad}^{\left(  2L\right)  }\left(  \lambda_{0}^{2}\right)  =3\lambda
_{0}^{2} \left[  I_{\log}^{\left(  2L\right)  }\left(  \lambda_{0}^{2}\right)
\right]  +\frac{i}{\left(  4\pi\right)  ^{2}}3\lambda_{0}^{2}\left[  I_{\log
}^{\left(  4\right)  }\left(  \lambda_{0}^{2}\right)  \right]  +C_{2}\ .
\end{equation}
Subtracting the two previous equations gives the scale relation for
$I_{quad}^{\left(  2L\right)  }\left(  \lambda^{2}\right)  $,
\begin{align}
I_{quad}^{\left(  2L\right)  }\left(  \lambda^{2}\right)   &  =I_{quad}%
^{\left(  2L\right)  }\left(  \lambda_{0}^{2}\right)  +3\left(  \lambda
^{2}-\lambda_{0}^{2}\right)  \left[  I_{\log}^{\left(  2L\right)  }\left(
\lambda_{0}^{2}\right)  \right]  +\frac{i}{\left(  4\pi\right)  ^{2}}3\left(
\lambda^{2}-\lambda_{0}^{2}\right)  \left[  I_{\log}^{\left(  4\right)
}\left(  \lambda_{0}^{2}\right)  \right] \nonumber\\
&  -\frac{i}{\left(  4\pi\right)  ^{2}}3\lambda^{2}\left[  I_{\log}^{\left(
4\right)  }\left(  \lambda_{0}^{2}\right)  -\frac{i}{\left(  4\pi\right)
^{2}}\frac{1}{2}\ln\left(  \frac{\lambda^{2}}{\lambda_{0}^{2}}\right)
\right]  \ln\left(  \frac{\lambda^{2}}{\lambda_{0}^{2}}\right)  \ .
\end{align}
With this scale relation one can change the physical mass $m$ inside the
undefined object $I_{quad}^{\left(  2L\right)  }$ in the sunrise diagram
$J_{SS}$, Eq.(\ref{Jss2}).

\section{Evaluating the integrals over Feynman parameters \label{sec_IOFP}}

In the evaluation of $J_{SS}$ (at end of Section (\ref{Sec_SS})), we left the
finite part of the result (\ref{Jss2}) as being represented by integrals over
Feynman parameters, which we now define as%
\begin{align}
\frac{\left(  4\pi\right)  ^{4}I_{F}}{m^{2}}  &  =\frac{k^{2}}{4}\int_{0}%
^{1}dz\ \left\{  1-\frac{2}{\left[  z\left(  1-z\right)  -1\right]  }%
+\frac{2\ln\left[  z\left(  1-z\right)  \right]  }{\left[  z\left(
1-z\right)  -1\right]  ^{2}}\right\} \nonumber\\
&  +\int_{0}^{1}dz\int_{0}^{1}dx\ \left[  \frac{3}{x}-k^{2}\left(  1-z\right)
\right]  \ \ln\left[  \frac{{p}^{2}z\left(  1-z\right)  +\left(  m^{2}-\mu
^{2}\right)  z-m^{2}}{\left(  m^{2}-\mu^{2}\right)  z-m^{2}}\right]  \ ,
\end{align}
where $k^{2}\equiv\frac{p^{2}}{m^{2}}$. Integration over $x$, in the second
integral above, followed by a trivial integrations-by-parts, gives an
one-dimensional integral representation which we write as
\begin{equation}
\frac{\left(  4\pi\right)  ^{4}I_{F}}{m^{2}}=I_{F}^{\left(  A\right)  }%
+I_{F}^{\left(  B\right)  },
\end{equation}%
\begin{align}
I_{F}^{\left(  A\right)  }  &  =-\frac{k^{2}}{4}+\frac{k^{2}}{4}\int_{0}%
^{1}dz\ \left\{  1-\frac{2}{\left[  z\left(  1-z\right)  -1\right]  }%
+\frac{2\ln\left[  z\left(  1-z\right)  \right]  }{\left[  z\left(
1-z\right)  -1\right]  ^{2}}\right\} \nonumber\\
&  \ -\int_{0}^{1}dz\ \frac{3-k^{2}\left(  1-z\right)  \left(  3z+1\right)
}{\sqrt{\left(  3z+1\right)  \left(  1-z\right)  }}\ln\left(  \frac
{3z+1+\sqrt{\left(  3z+1\right)  \left(  1-z\right)  }}{3z+1-\sqrt{\left(
3z+1\right)  \left(  1-z\right)  }}\right)  \ ,
\end{align}%
\begin{equation}
I_{F}^{\left(  B\right)  }=-\frac{1}{2}\int_{0}^{1}dz\ \frac{k^{2}\left(
k^{2}+3\right)  z^{2}-\left(  k^{4}-4k^{2}+3\right)  z+\left(  k^{2}-5\right)
}{k^{2}\sqrt{P_{4}}}\ln\left(  \frac{\left(  z-z_{2}\right)  \left(
z-z_{3}\right)  -\sqrt{P_{4}}}{\left(  z-z_{2}\right)  \left(  z-z_{3}\right)
+\sqrt{P_{4}}}\right)  \ , \label{I_F_B}%
\end{equation}
where $P_{4}=\left(  z-z_{1}\right)  \left(  z-z_{2}\right)  \left(
z-z_{3}\right)  \left(  z-z_{4}\right)  $ is a quartic polynomial and $z_{i}$
are its roots,%
\begin{equation}
\left\{
\begin{array}
[c]{c}%
z_{1}=\frac{1}{k^{2}}\ ,\ \ z_{2}=\frac{\left(  k^{2}-3\right)  }{2k^{2}%
}-\frac{\sqrt{k^{4}-10k^{2}+9}}{2k^{2}},\\
\ z_{3}=\frac{\left(  k^{2}-3\right)  }{2k^{2}}+\frac{\sqrt{k^{4}-10k^{2}+9}%
}{2k^{2}},\ \ z_{4}=1\ .
\end{array}
\right.
\end{equation}
The integral $I_{F}^{\left(  A\right)  }$ can be easily performed and written
in terms of ordinary MPLs, yielding
\begin{equation}
I_{F}^{\left(  A\right)  }=\frac{k^{2}}{2}+\left(  \xi_{+}-\xi_{-}\right)
\left\{  G\left(  0,\frac{1}{\xi_{-}},1\right)  -G\left(  0,\frac{1}{\xi_{+}%
},1\right)  \right\}  \ ,
\end{equation}
with the standard definition for the MPLs%
\begin{equation}
G\left(  a_{1},...,a_{n};x\right)  =\int_{0}^{x}\frac{dx^{\prime}}{x^{\prime
}-a_{1}}G\left(  a_{2},...,a_{n};x^{\prime}\right)  ~,
\end{equation}
and $\xi_{\pm}=\frac{1\pm i\sqrt{3}}{2}$.

In its turn, the integral $I_{F}^{\left(  B\right)  }$ cannot be written in
terms of these ordinary MPLs. This is due to the fact that the integral
involved is of elliptic type. In order to circumvent this problem, in this
last decade an elliptic generalization of the MPLs (named of eMPLs) was
proposed and developed, in a slight different forms, in a series of papers.
The aim of this section is to compute $I_{F}^{\left(  B\right)  }$ and express
the result in terms of such eMPLs. Here we follow closely some of the
notations and definitions stated in Refs. \cite{Broedel, Broedel2}, where the
authors define the eMPLs ($E_{4}$) as iterated integrals,%
\begin{equation}
E_{4}\left(
\genfrac{}{}{0pt}{}{n_{1}\ldots n_{k}}{c_{1}\ldots c_{k}}%
;z\right)  =\int_{0}^{z}dz^{\prime}\psi_{n_{1}}\left(  c_{1},z^{\prime
}\right)  E_{4}\left(
\genfrac{}{}{0pt}{}{n_{2}\ldots n_{k}}{c_{2}\ldots c_{k}}%
;z^{\prime}\right)  \ ,
\end{equation}
with the suitable integration kernels $\psi_{n_{i}}\left(  c_{i},z\right)  $
defined by%
\begin{align*}
\psi_{0}\left(  0,z\right)   &  =\frac{\sqrt{\left(  z_{1}-z_{3}\right)
\left(  z_{2}-z_{4}\right)  }}{2\sqrt{P_{4}}}\ ,\ \ \ \ \ \ \ \psi_{1}\left(
0,z\right)  =\frac{1}{z}\ ,\\
\psi_{-1}\left(  0,z\right)   &  =\frac{z_{1}z_{4}}{z\sqrt{P_{4}}}-\frac{1}%
{z}\ ,\ \ \ \ \ \ \psi_{-1}\left(  \infty,z\right)  =\frac{z}{\sqrt{P_{4}}%
}\ ,\\
\psi_{-2}\left(  \infty,z\right)   &  =\frac{z}{\sqrt{P_{4}}}Z_{4}\left(
z\right)  -\frac{2}{\sqrt{\left(  z_{1}-z_{3}\right)  \left(  z_{2}%
-z_{4}\right)  }}\ ,\\
\psi_{-2}\left(  0,z\right)   &  =\frac{z_{1}z_{4}}{z\sqrt{P_{4}}}Z_{4}\left(
z\right)  \ .
\end{align*}
In the kernels defined above we have
\begin{equation}
Z_{4}\left(  z\right)  =\int_{z_{1}}^{z}dz\ \left\{  \widetilde{\Phi}%
_{4}\left(  z\right)  +2\sqrt{\left(  z_{1}-z_{3}\right)  \left(  z_{2}%
-z_{4}\right)  }\left(  \frac{E\left(  \lambda\right)  }{K\left(
\lambda\right)  }-\frac{\left(  2-\lambda\right)  }{3}\right)  \frac{1}%
{\sqrt{P_{4}}}\right\}  \ ,
\end{equation}%
\begin{align}
\widetilde{\Phi}_{4}\left(  z\right)   &  =\frac{2}{\sqrt{\left(  z_{1}%
-z_{3}\right)  \left(  z_{2}-z_{4}\right)  }\sqrt{P_{4}}}\left\{  z^{2}%
-\frac{1}{2}\left(  z_{1}+z_{2}+z_{3}+z_{4}\right)  z\right. \nonumber\\
&  \left.  +\frac{1}{6}\left(  z_{1}z_{2}+z_{1}z_{3}+z_{1}z_{4}+z_{2}%
z_{3}+z_{2}z_{4}+z_{3}z_{4}\right)  \right\}  \ ,
\end{align}
with $K\left(  \lambda\right)  $ and $E\left(  \lambda\right)  $ being two
well-known elliptic integrals,
\begin{align}
K\left(  \lambda\right)   &  =\int_{0}^{1}dt\ \frac{1}{\sqrt{\left(
1-t^{2}\right)  \left(  1-\lambda t^{2}\right)  }}\ ,\\
E\left(  \lambda\right)   &  =\int_{0}^{1}dt\ \frac{1-\lambda t^{2}}%
{\sqrt{\left(  1-t^{2}\right)  \left(  1-\lambda t^{2}\right)  }}\ ,\\
\lambda &  =\frac{\left(  z_{1}-z_{4}\right)  \left(  z_{2}-z_{3}\right)
}{\left(  z_{1}-z_{3}\right)  \left(  z_{2}-z_{4}\right)  }\ .
\end{align}
If we adopt the following integral representation for the logarithm showed in
Eq. (\ref{I_F_B}), as suggested in \cite{Broedel2}, i.e.,%
\begin{equation}
\ln\left(  \frac{\left(  z-z_{2}\right)  \left(  z-z_{3}\right)  -\sqrt{P_{4}%
}}{\left(  z-z_{2}\right)  \left(  z-z_{3}\right)  +\sqrt{P_{4}}}\right)
=E_{4}\left(
\genfrac{}{}{0pt}{}{-1}{\infty}%
;z\right)  -E_{4}\left(
\genfrac{}{}{0pt}{}{-1}{0}%
;z\right)  -E_{4}\left(
\genfrac{}{}{0pt}{}{1}{0}%
;z\right)  \ ,
\end{equation}
we obtain, after some algebraic manipulations,%
\begin{align}
I_{3}^{\left(  B\right)  }  &  =\frac{\left(  k^{2}+3\right)  }{2}\nonumber\\
&  +\mathcal{C}_{1}\left\{  E_{4}\left(
\genfrac{}{}{0pt}{}{-2}{\infty}%
;1\right)  -E_{4}\left(
\genfrac{}{}{0pt}{}{-2}{0}%
;1\right)  -Z_{4}\left(  1\right)  \left[  E_{4}\left(
\genfrac{}{}{0pt}{}{-1}{\infty}%
;1\right)  -E_{4}\left(
\genfrac{}{}{0pt}{}{-1}{0}%
;1\right)  \right]  \right\} \nonumber\\
&  +\mathcal{C}_{2}\left\{  E_{4}\left(
\genfrac{}{}{0pt}{}{-1}{\infty}%
\genfrac{}{}{0pt}{}{-1}{\infty}%
;1\right)  -E_{4}\left(
\genfrac{}{}{0pt}{}{-1}{\infty}%
\genfrac{}{}{0pt}{}{-1}{0}%
;1\right)  +E_{4}\left(
\genfrac{}{}{0pt}{}{1}{0}%
\genfrac{}{}{0pt}{}{-1}{\infty}%
;1\right)  \right\} \nonumber\\
&  +\mathcal{C}_{3}\left\{  E_{4}\left(
\genfrac{}{}{0pt}{}{0}{0}%
\genfrac{}{}{0pt}{}{-1}{\infty}%
;1\right)  -E_{4}\left(
\genfrac{}{}{0pt}{}{0}{0}%
\genfrac{}{}{0pt}{}{-1}{0}%
;1\right)  +E_{4}\left(
\genfrac{}{}{0pt}{}{1}{0}%
\genfrac{}{}{0pt}{}{0}{0}%
;1\right)  \right\}  \ ,
\end{align}
with the coefficients%
\begin{align}
\mathcal{C}_{1}  &  =\frac{\left(  k^{2}+3\right)  \sqrt{\left(  z_{1}%
-z_{3}\right)  \left(  z_{2}-z_{4}\right)  }}{4}\ ,\\
\mathcal{C}_{2}  &  =\frac{\left(  k^{4}-4k^{2}+3\right)  -\left(
k^{2}+3\right)  \left(  k^{2}-1\right)  }{2k^{2}}\ ,\\
\mathcal{C}_{3}  &  =\frac{\left(  k^{2}+3\right)  \left(  k^{4}-3\right)
-6k^{2}\left(  k^{2}-5\right)  }{6k^{4}\sqrt{\left(  z_{1}-z_{3}\right)
\left(  z_{2}-z_{4}\right)  }}\nonumber\\
&  +\frac{\left(  k^{2}+3\right)  }{2}\left(  \frac{E\left(  \lambda\right)
}{K\left(  \lambda\right)  }-\frac{\left(  2-\lambda\right)  }{3}\right)
\sqrt{\left(  z_{1}-z_{3}\right)  \left(  z_{2}-z_{4}\right)  }\ .
\end{align}
In this way, the formula obtained for $J_{SS}$ in terms of eMPLs is very
simple. In order to complete the analysis of the $J_{SS}$ diagram, in the next
section we will show the numerical analysis of the obtained results.

\section{The numerical analysis}

Let us complete our study about the sunrise diagram by performing a brief
numerical analysis of the obtained result. With respect to the finite part
obtained for $J_{SS}$ in Eq. (\ref{Jss2}), the double integral over Feynman
parameters
\begin{equation}
F=\int_{0}^{1}dz\int_{0}^{1}dx\ \left[  \frac{3}{k^{2}}\frac{1}{x}-\left(
1-z\right)  \right]  \ \ln\left[  \frac{\left[  k^{2}z\left(  1-z\right)
+z-1\right]  x\left(  1-x\right)  -z}{\left(  z-1\right)  x\left(  1-x\right)
-z}\right]  \ ,
\end{equation}
deserves more attention. For numerical calculations it is convenient to write
(after integration in $x$)
\begin{equation}
F=\int_{0}^{1}dz\ \left\{  -\frac{3}{2{k}^{2}}\ln^{2}\left(  \frac{1+\sqrt{X}%
}{1-\sqrt{X}}\right)  +\frac{\left(  1-z\right)  }{\sqrt{X}}\ln\left(
\frac{1+\sqrt{X}}{1-\sqrt{X}}\right)  \right\}  -\left\{  X\rightarrow
X_{0}\right\}  \ , \label{F_1}%
\end{equation}
with%
\begin{equation}
X=\frac{\left(  z-z_{1}\right)  \left(  z-z_{4}\right)  }{\left(
z-z_{2}\right)  \left(  z-z_{3}\right)  },\ \ \ X_{0}=\frac{1-z}{1+3z}\ .
\end{equation}
The second integral, involving the variable $X_{0}$, is straightforward. Let
us focus on the first integral, which we from now on will denominate
$\widetilde{F}$. In the following, we split up our analysis into three
possible regions of $k^{2}$.

\begin{figure}[h]
\centering
\includegraphics[width=230pt,height=230pt]{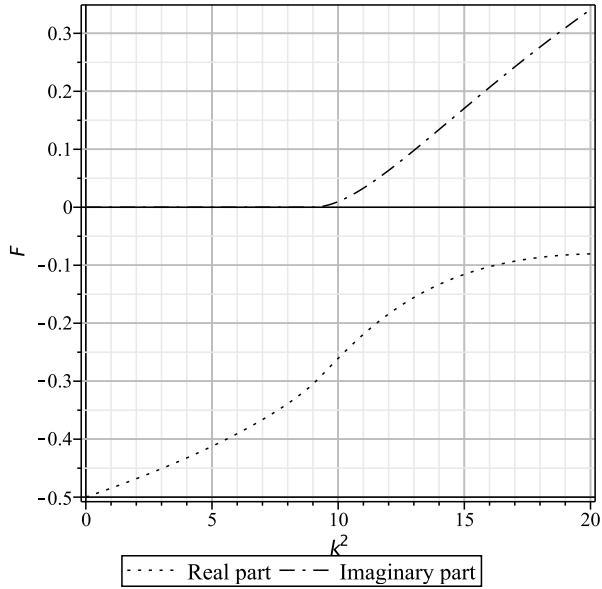} \caption{Representation
of the function $F$.}%
\label{plot}%
\end{figure}

If $k^{2}\leq1$, we have $0\leq X\leq1$ for the full integration interval
($0\leq z\leq1$), which makes the numerical integration of (\ref{F_1}) easily.
In the interval $1<k^{2}<9$, the roots $z_{2}$ and $z_{3}$ are complex
conjugate of each other. In this case, it is convenient to split the interval
of integration into two parts, since
\[
\left\{
\begin{array}
[c]{c}%
0\leq X\leq1\text{ for }0\leq z\leq z_{1}\ ,\\
X\leq0\text{ for }z_{1}\leq z\leq1\ .
\end{array}
\right.
\]
Then, we find out that numerical integration of $\widetilde{F}$, within the
interval $1<k^{2}<9$, is most easily performed if written in the form%
\begin{align}
\widetilde{F}\left(  1<k^{2}<9\right)  =  &  \int_{0}^{z_{1}}dz\ \left\{
-\frac{3}{2{k}^{2}}\ln^{2}\left(  \frac{1+\sqrt{X}}{1-\sqrt{X}}\right)
+\frac{\left(  1-z\right)  }{\sqrt{X}}\ln\left(  \frac{1+\sqrt{X}}{1-\sqrt{X}%
}\right)  \right\} \nonumber\\
&  +\int_{z_{1}}^{1}dz\ \left\{  \frac{6}{{k}^{2}}\arctan^{2}\left(
\sqrt{\left\vert X\right\vert }\right)  +\frac{2\left(  1-z\right)  }%
{\sqrt{\left\vert X\right\vert }}\arctan\left(  \sqrt{\left\vert X\right\vert
}\right)  \right\}  \ .
\end{align}
In its turn, for $k^{2}\geq9$ (the three massive particle cut) there is a
threshold at $k^{2}=9$ and all the roots are real and ordered $z_{1}%
<z_{2}<z_{3}<z_{4}$. In this region we find%
\[
\left\{
\begin{array}
[c]{c}%
0\leq X\leq1\ \text{for }0\leq z\leq z_{1}\ ,\\
X\leq0\text{ for }z_{1}\leq z\leq z_{2}\ \text{and }z_{3}\leq z\leq1\ ,\\
X>1\text{ for }z_{2}\leq z\leq z_{3}\ .
\end{array}
\right.
\]
For numerical calculations, we found it convenient split out this integration
region into the above intervals, which gives%
\begin{align}
\widetilde{F}\left(  k^{2}\geq9\right)   &  =3\pi^{2}\frac{\sqrt{\left(
k^{2}-9\right)  \left(  k^{2}-1\right)  }}{2k^{4}}\nonumber\\
&  -\int_{0}^{z_{1}}dz\ \left\{  \frac{3}{2{k}^{2}}\ln^{2}\left(
\frac{1+\sqrt{X}}{1-\sqrt{X}}\right)  -\frac{\left(  1-z\right)  }{\sqrt{X}%
}\ln\left(  \frac{1+\sqrt{X}}{1-\sqrt{X}}\right)  \right\} \nonumber\\
&  +\int_{z_{1}}^{z_{2}}dz\ \left\{  \frac{6}{{k}^{2}}\arctan^{2}\left(
\sqrt{\left\vert X\right\vert }\right)  +\frac{2\left(  1-z\right)  }%
{\sqrt{\left\vert X\right\vert }}\arctan\left(  \sqrt{\left\vert X\right\vert
}\right)  \right\} \nonumber\\
&  -\int_{z_{2}}^{z_{3}}dz\ \left\{  \frac{3}{2{k}^{2}}\ln^{2}\left(
\frac{1+\frac{1}{\sqrt{X}}}{1-\frac{1}{\sqrt{X}}}\right)  -\frac{\left(
1-z\right)  }{\sqrt{X}}\ln\left(  \frac{1+\frac{1}{\sqrt{X}}}{1-\frac{1}%
{\sqrt{X}}}\right)  \right\} \nonumber\\
&  +\int_{z_{3}}^{1}dz\ \left\{  \frac{6}{{k}^{2}}\arctan^{2}\left(
\sqrt{\left\vert X\right\vert }\right)  +\frac{2\left(  1-z\right)  }%
{\sqrt{\left\vert X\right\vert }}\arctan\left(  \sqrt{\left\vert X\right\vert
}\right)  \right\} \nonumber\\
&  +i\pi\Theta\left(  k^{2}-9\right)  \int_{z_{2}}^{z_{3}}dz\ \left\{
\frac{\left(  1-z\right)  }{\sqrt{X}}-\frac{3}{{k}^{2}}\ln\left(
\frac{1+\frac{1}{\sqrt{X}}}{1-\frac{1}{\sqrt{X}}}\right)  \right\}  \ .
\end{align}
With the formulae presented above, one can easily build up the graph of $F$ as
a function of $k^{2}$ by using some standard numerical package to perform the
integrals. It is showed in the Fig. (\ref{plot}).

\section{Final remarks and conclusions}

In the present work we considered in details the explicit calculation of the
well-known two-loop equal mass sunrise Feynman diagram. The investigation was
based mainly on the ideas developed in the alternative method described in
Section \ref{sec_method}. This referred strategy gives us a consistent
framework to promote investigations in perturbative calculations in situations
where the traditional methods are not consistent or not applicable. It does
not have limitations of applicability since it works equally well in even and
odd space-time dimensions, in the presence of tensors and pseudo-tensors
amplitudes in even dimensions and in the context of renormalizable and
nonrenormalizable theories. In particular, in all situations where the DR
applies, one could find, in principle, a map that puts the results in a
precise correspondence. This represents an obvious advantage in calculations
involving multi-loops, since that almost all results available on such issue
are performed within the DR context and would be desirable to have at our
disposal an alternative approach in order to be possible to compare the
results. Many previously investigations made revealed the general and
consistent character of this approach when applied to one-loop calculations.
In the present paper we have shown that this method can be equally applicable
for perturbative calculations involving two-loop graphs, without need to add
any new rule, when compared with one-loop calculations. The procedure is
simple to apply, consisting in rewrite the integrand of divergent Feynman
integrals, by means of an alternative representation of propagators, into a
sum of terms splitted into two distinct classes.

The first of such classes is formed by terms which have denominators that are
momenta and masses dependent, but are finite when integrated over the loop
momenta. Therefore, these finite integrals are ready to be integrated over the
loop momenta and the results are written through integrals over Feynman
parameters. In the Section \ref{Sec_SS} we showed that, for the sunrise graph
calculation, this class is represented by the following terms (see eq.
(\ref{Jss2}))
\begin{align}
&  \frac{1}{\left(  4\pi\right)  ^{4}}\frac{p^{2}}{4}\int_{0}^{1}dx\ \left\{
1-\frac{2}{\left[  x\left(  1-x\right)  -1\right]  }+\frac{2\ln\left[
x\left(  1-x\right)  \right]  }{\left[  x\left(  1-x\right)  -1\right]  ^{2}%
}\right\} \nonumber\\
&  +\frac{1}{\left(  4\pi\right)  ^{4}}\int_{0}^{1}dy\int_{0}^{1}dx\ \left[
\frac{3m^{2}}{x}-p^{2}\left(  1-y\right)  \right]  \ \ln\left\{  \frac{{p}%
^{2}y\left(  1-y\right)  +\left(  m^{2}-\mu^{2}\right)  y-m^{2}}{\left(
m^{2}-\mu^{2}\right)  y-m^{2}}\right\}  \ .
\end{align}
The first integral above can be done easily and the second one is more
involving since it is of elliptic type, which is not a surprise at all. In the
Section \ref{sec_IOFP}, we in particular provide the explicit representation
of this integral in terms of eMPLs functions, which are, in some sense,
generalizations of the well-known MPLs functions. The expression obtained has
a simple algebraic form.

The second class contain divergent integrals which are momenta and masses
(physical) independent. Such terms are not integrated out, since they are
undefined quantities. Instead, they are organized through (tensor) surface
terms and scalar basic divergent objects. The calculation of $J_{SS}$ revealed
that this class are composed by%
\begin{equation}
\left[  I_{quad}^{\left(  2L\right)  }\left(  m^{2}\right)  \right]  -\frac
{i}{32\pi^{2}}\left\{  p^{2}\left[  I_{\log}^{\left(  4\right)  }\left(
m^{2}\right)  \right]  +p^{\mu}p^{\nu}\left[  \Delta_{\mu\nu}^{\left(
4\right)  }\left(  m^{2}\right)  \right]  \right\}  \ ,
\end{equation}
where $I_{\log}^{\left(  4\right)  }$ and $\Delta_{\mu\nu}^{\left(  4\right)
}$ are typical quantities of one-loop integrals and $I_{quad}^{\left(
2L\right)  }$ is a object which is present in two-loop calculations. One can
note that, in the adopted strategy, such divergent quantities emerge having
coefficients which are polynomial in the external momentum and, therefore,
they could be naturally removed in a renormalization process without the
requirement of evaluating the integral over the internal momenta, which is
ill-defined. In this sense, one can say that no regularization is need in practice.

The procedure applied in this work can also be used to calculate others
topologies of two-loop graphs or yet graphs of higher order in the
perturbative expansion, with no restriction of applicability. Since it has
simple and universal rules, independent of Feynman integral considered, its
implementation is easy and systematic. Thus, it can be considered, at least,
an alternative consistent method to DR for treat divergent multi-loop graph.
Beside that, many techniques invented to deal with dimensionally regulated
integrals can also be used together with the method, as we have given an
example in Section \ref{Sec_SS}, where we used the technique of
integration-by-parts in order to obtain the identities (\ref{TD1}) and
(\ref{TD2}).

\end{document}